\documentstyle[12pt]{article}

\begin{document}

\begin{center}
\large{{\bf Dirac Constraint Quantization of a Dilatonic Model of
Gravitational Collapse}}
\end{center}

\begin{center}
Karel V.~Kucha\v{r}\\
Department of Physics, University of Utah, Salt Lake City, UT 84112
\end{center}

\begin{center}
Joseph D.~Romano\\
Department of Physics, University of Wisconsin-Milwaukee, Milwaukee \\ 
WI 53201
\end{center}

\begin{center}
Madhavan Varadarajan\\
Department of Physics, University of Utah, Salt Lake City, UT 84112
\end{center}

\section*{Abstract}
We present an anomaly-free Dirac constraint quantization of the 
string-inspired dilatonic gravity (the CGHS model) in an open 
2-dimensional spacetime.
We show that the quantum theory has the same degrees of freedom as 
the classical theory; namely, all the modes of the scalar field on an 
auxiliary flat background, supplemented by a single additional variable 
corresponding to the primordial component of the black hole mass. 
The functional Heisenberg equations of motion for these dynamical 
variables and their canonical conjugates are linear, and they have 
exactly the same form as the corresponding classical equations. 
A canonical transformation brings us back to the physical geometry 
and induces its quantization.

\bigskip
\noindent PACS number(s): 04.60.Kz, 04.60.Ds, 04.70.Dy, 04.20.Ha, 
11.10.Kk

\newpage
\section*{1. Introduction}

The formation of black holes by collapsing matter fields and their
subsequent Hawking evaporation are well understood only within the 
semiclassical approximation.
Unfortunately, an analysis of the final stages of black hole 
evaporation, its possible remnants, the information problem, and the 
fate of a final singularity, requires quantum gravity. 
In its absence, the best one can do is to turn to simplified 
models of gravitational collapse.

The least idealized of such models is obtained by a dimensional 
reduction of spacetime geometry coupled to a collapsing massless scalar 
field under the assumption of spherical symmetry. 
This was set up by Berger, Chitre, Moncrief, and Nutku (BCMN) \cite{bcmn} 
in 1972, and later corrected by Unruh \cite{un}.
However, even this simple model is not classically exactly solvable, 
and its quantization remains problematic.

By what formally appears to be a minor modification of the BCMN action,
Callan, Giddings, Harvey, and Strominger (CGHS) \cite{cghs} turned the BCMN
model into one whose general classical solution is explicitly known. 
Their choice of the action was motivated by the effective action describing 
spherical modes of extremal dilatonic black holes in four or higher 
dimensions \cite{hor} and the spacetime action for noncritical strings
\cite{witt}. 
Recalling its origin and laudibly avoiding the acronym, the CGHS model is 
often referred to as `string-inspired dilatonic gravity'.
It is viewed as a genuine 2-dimensional theory.
 
Surprisingly, the quantization of even this classically exactly solvable 
model is far from trivial. 
It has been discussed from different points of view by numerous 
investigators \cite{str}. 
The aim of this paper is to show that the canonical Dirac constraint 
quantization of the dilatonic model admits an exact solution of its quantum 
dynamics such that its physical degrees of freedom exactly correspond to 
those of the classical theory.

Canonical quantization of the string-inspired dilatonic gravity has been 
previously studied by Mikovi\'{c} \cite{mik 1,mik 3,mik 4} and by 
Jackiw and his collaborators \cite{jac 1,jac 2,jac 3,jac 4}.\footnote{Put 
into a general setting of Poisson $\sigma $-models, it got a lucid treatment 
by Strobl, Kl\"{o}sch, and Schaller \cite{strob}. 
Primordial (matter-free) dilatonic black holes are canonically quantized 
in \cite{kun}. 
Other relevant references are \cite{oth 1}.}

Mikovi\'{c} quantized the model after the ADM (Arnowitt, Deser, and Misner
\cite{adm}) reduction to a constant extrinsic time foliation. 
He focused on the clarification of issues connected with the generation of 
Hawking radiation and unitarity of quantum black hole evolution. 
By fixing the foliation, however, he sidestepped the problems of possible 
non-integrability of the functional Schr\"{o}dinger equation caused by 
anomalies.

In a series of carefully written papers, Jackiw and his coworkers discussed 
in detail different constraint quantization techniques (in particular the 
functional Schr\"{o}dinger quantization and the BRST 
(Becchi-Rouet-Stora-Tyutin) quantization) and the discrepancies of the 
resulting quantum theories. 
In \cite{jac 3}, they concluded that the functional Schr\"{o}dinger 
equation approach seems unable to produce a large (or infinite) number of 
physical states, while the BRST approach leads to an infinite number of 
states corresponding to a single complete set of oscillators complemented 
by two homogeneous modes.
More recently \cite{jac 4}, they noticed the existence of a complete set 
of oscillator states in the functional Schr\"{o}dinger equation approach, 
which reduces the mentioned discrepancy to an extra homogeneous mode in the 
BRST approach.

In the above papers, the explicit handling of the quantum model is made 
possible by a series of transformations from the original CGHS variables. 
The first of these is a rescaling of the physical 2-dimensional metric 
into an auxiliary flat metric by the dilaton \cite{jac 3}. 
The second is a canonical transformation that casts the constraints into 
those of a bosonic string in a 3-dimensional target space \cite{jac 2}. 
(This form is the starting point for the BRST quantization.) 
A long time ago, one of us showed \cite{kuch 1} that this type of
constraints can be further simplified by a third canonical transformation 
that brings the constraints into the form appropriate for the parametrized 
theory \cite{par} of a single scalar field propagating on a 2-dimensional 
Minkowskian background. 
(This was the starting point for a functional Schr\"{o}dinger equation 
treatment of the bosonic string \cite{torre}.) 
The same approach was adopted in \cite{jac 3} to quantize the dilatonic 
model.

But by performing the canonical transformation to the would-be embedding 
variables in a series of steps, one can lose sight of their basic geometric 
significance. 
We present an alternative derivation of a transformation that ensures that 
our embedding variables $X^{\pm}(x)$  have the proper physical interpretation 
on the physical spacetime manifold equipped with the double null Minkowskian 
coordinates $X^{\pm}$ of its auxiliary flat metric (section 2c). 
This enables us to maintain the physical interpretation of the functional 
Heisenberg equations of motion (or the equivalent functional Schr\"{o}dinger 
equation) in the quantum theory.

Still, even the Dirac constraint quantization of a parametrized scalar
field is not straightforward. 
Commutators of the constraint operators develop an anomaly, which leads to 
inconsistencies when one imposes the constraints as operator restrictions 
on physical states. 
One of us faced this problem earlier \cite{kuch 2} when studying the Dirac 
constraint quantization of a massless scalar field propagating on a flat 
Minkowskian cylinder ${\rm I\!R}\times S^1$. 
There it was shown how to remove the anomaly by an embedding-dependent 
factor ordering of the constraints. 
An analogous modification of the constraints was used in \cite{jac 3,jac 4} 
to quantize the dilatonic model.
But unfortunately, the change of topology from 
${\rm I\!R}\times{\rm I\!R}$ to ${\rm I\!R}\times S^1$ is in conflict 
with the original geometric interpretation of dilatonic gravity as a theory 
of black hole formation in the physical curved spacetime 
(see the Appendix).\footnote{How to put the CGHS model on 
${\rm I\!R}\times S^1$ is discussed in \cite{mik 4}.
A thorough study of possible ${\rm I\!R}\times S^1$ compactifications 
in different versions of dilatonic gravity and their consequences (like 
geodesic incompleteness or the presence of closed timelike curves) was 
undertaken in \cite{strob}.} 
For this reason, we pay close attention to the open-space boundary 
conditions in dilatonic gravity, and point out that only they enable us to 
turn the transformation between the geometric and embedding variables into 
a truly bona fide, one-to-one, canonical transformation (section 3). 
This leads us finally to a consistent Dirac constraint quantization of the 
original dilatonic model (section 4). 
The removal of the anomaly is best understood by passing to the Heisenberg 
picture (sections 2d and 4a,b) in which the constraints are manifestly 
anomaly-free.\footnote{This conceptual strategy is clarified for a 
finite-dimensional parametrized system in \cite{kuch 3}.} 
The quantum dynamics of the dilatonic model is entirely explicit in this 
picture, but it can also be recast into the Schr\"{o}dinger picture, which
is traditionally associated with the Dirac constraint quantization 
(sections 4c,d,e). 
We conclude that the quantum theory of the string-inspired dilatonic 
gravity has exactly the same degrees of freedom as the classical theory;
namely, all the modes of the scalar field on the auxiliary open flat 
background, supplemented by a single additional degree of freedom 
corresponding to the primordial component of the black hole mass. 
This conclusion is in general agreement with the ${\rm I\!R}\times S^1$ 
results obtained in \cite{jac 4}. 
Moreover, the functional Heisenberg equations of motion for these dynamical 
variables and their canonical conjugates are linear, and they have exactly 
the same form as the corresponding classical equations.

One can return by the canonical transformation from these dynamical 
variables back to the original geometric variables. 
This enables us in the end to pose some relevant questions about the quantum 
physical geometry of dilatonic gravity (section 5). 

\subsection*{Notation} 
Besides standard conventions, we will use the following notation 
throughout this paper: 
Arguments of functions will be enclosed in round brackets (e.g., $y(X)$),
while arguments of function{\it{al}}s will be enclosed in square
brackets (e.g., $S[y,\gamma_{\alpha\beta},f]$).
If a quantity is simultaneously a function of some variables, say $x$, 
and a functional of other variables, say $X$, we will use both round and 
square brackets as in $n^\alpha(x;X]$, with the semi-colon separating 
the function and functional dependence.
In the double null coordinates $X^\alpha=(X^+,X^-)$, many quantities 
depend only on $X^+$ or $X^-$, but not on both variables.
We will emphasize this by using only $X^+$ or $X^-$ as an argument of 
that function or functional.
For example, while $f(X)$ means that $f$ is a function of both $X^+$ 
and $X^-$, $f_{,+}(X^+)$ and $f_{,-}(X^-)$ mean that the derivatives
$f_{,+}$ and $f_{,-}$ depend only on $X^+$ and $X^-$, respectively.
Moreover, $f_{,\pm}(X^\pm)$ will serve as a shorthand notation to 
denote the function dependence of both $f_{,+}$ and $f_{,-}$ simultaneously.
Finally, $A_\pm(x;X]$ means that $A_+$ and $A_-$ are 
functions of $x$ and functionals of both $X^+$ and $X^-$.
This is to be contrasted with the functional dependence of $h_+$ and
$h_-$ as indicated by $h_\pm(x;X^\pm,f,\pi_f]$.

\section*{2. Classical theory}

\subsection*{2a. Spacetime action and equations of motion}

We take, as our starting point, the spacetime action for dilatonic
gravity written in the form
\begin{equation}
S[y,\gamma_{\alpha\beta},f]={1\over 2}\int d^2\!X\ |\gamma|^{{1\over 2}}
\left(y R[\gamma]+4\kappa^2-\gamma^{\alpha\beta}f_{,\alpha}f_{,\beta}\right)
\ .
\label{eq:S}
\end{equation}
Here $y$ is the dilaton field, $\gamma_{\alpha\beta}$ is the spacetime 
metric (signature $(-+)$), and $f$ is a conformally coupled scalar field.
$R[\gamma]$ denotes the scalar curvature of $\gamma_{\alpha\beta}$, and 
$\kappa$ is a constant having the dimensions of inverse 
length.\footnote{In special relativity, 
$c=1$ and the basic dimensions are mass, $M$, and length, $L$. 
Action has the dimension $[S]=ML$. 
We let the metric $\gamma_{\alpha \beta}$ and the dilaton $y$ be 
dimensionless; the spatial dimension $L$ is carried by the coordinates $X$. 
To match the dimension of $R[\gamma ]$, the classical constant $\kappa$ 
must have the dimension $[\kappa]= L^{-1}$. 
The scalar field with dimension $[f]=M^{1\over 2}L^{1\over 2}$ yields the 
correct dimensionality for the matter action. 
Matter is coupled to gravity by Newton's constant $G$, which in a 2d
spacetime has the dimension $[G]= M^{-1}L^{-1}$. 
The coupling constant $G^{-1}$ in front of the dilatonic part of the 
action (\ref{eq:S}) restores its proper dimensionality. 
By setting $G=1$, we agree to measure mass in units of $[M] = L^{-1}$. 
By the same decision, the action, which is measured in the units of 
$G^{-1}$, and the scalar field measured in the units of $G^{-{1\over 2}}$, 
become dimensionless. 
The only remaining basic dimension is $L$. 
None of these dimensional considerations has anything to do with quantum 
theory.}
To interpret the theory, we will treat $\gamma_{\alpha\beta}$ as an 
auxiliary metric and 
\begin{equation}
\bar\gamma_{\alpha\beta}:=y^{-1}\gamma_{\alpha\beta}
\label{eq:physical_metric}
\end{equation}
as the physical ``black hole'' metric.
The action (\ref{eq:S}) is obtained from the original CGHS action (which 
is a functional of the physical metric $\bar {\gamma}_{\alpha \beta}$) by 
the rescaling (\ref{eq:physical_metric}).\footnote{If we put 
$\lambda=4\kappa^2$, $\eta=y$, and change the signature of 
$\gamma_{\alpha\beta}$, we recover the dilatonic action given in \cite{jac 3}.
If we put $\lambda=\kappa$, $\phi=-{\scriptstyle{1\over 2}}\ln y$, and 
$\bar\gamma_{\alpha\beta}=y^{-1}\gamma_{\alpha\beta}$, we recover 
(up to a boundary term) the dilatonic action given in \cite{cghs}.}

The equations of motion of dilatonic gravity are derived, as usual, 
by varying the spacetime action (\ref{eq:S}) with respect to all of its
arguments.
(See also \cite{jac 3} and \cite{cghs}.)
Variation of $y$ implies
\begin{equation}
R[\gamma]=0\ .
\label{eq:1}
\end{equation}
Variation of $f$ implies
\begin{equation}
\Box_\gamma f:=\left(|\gamma|^{{1\over 2}}\gamma^{\alpha\beta}
f_{,\beta}\right){}_{,\alpha}=0\ .
\label{eq:2}
\end{equation}
Variation of the (contravariant) spacetime metric $\gamma^{\alpha\beta}$ 
implies
\begin{equation}
y_{;\alpha\beta}-\gamma_{\alpha\beta}\Box_\gamma y=
-\left(T_{\alpha\beta}+2\kappa^2\gamma_{\alpha\beta}\right)\ ,
\label{eq:yeqn}
\end{equation}
where
\begin{equation}
T_{\alpha\beta}:=f_{,\alpha}f_{,\beta}-{1\over 2}
\gamma_{\alpha\beta}\gamma^{\mu\nu}f_{,\mu}f_{,\nu}
\label{eq:energy-momentum}
\end{equation}
is the energy-momentum tensor of the scalar field $f$.
Equation (\ref{eq:yeqn}) can be simplified by contracting it 
with $\gamma^{\alpha\beta}$.
This yields
\begin{equation}
\Box_\gamma y = 4\kappa^2\ .
\label{eq:boxy}
\end{equation}
Substituting (\ref{eq:boxy}) back into (\ref{eq:yeqn}), we find
\begin{equation}
y_{;\alpha\beta}=
-\left(T_{\alpha\beta}-2\kappa^2\gamma_{\alpha\beta}\right)\ .
\label{eq:3}
\end{equation}

We solve the equations of motion (\ref{eq:1}), (\ref{eq:2}), (\ref{eq:3}) 
as follows:
In 2-dimensions, $R[\gamma]=0$ implies that spacetime is flat.
Thus, we can introduce Minkowskian coordinates $(T,Z)$, or the equivalent 
double-null coordinates
\begin{equation}
X^\pm:=Z\pm T\ , \label{eq:9}
\end{equation}
for which the spacetime line elements takes the form
\begin{equation}
ds^2=dX^+ dX^-\ .
\end{equation}
The solution of the wave equation (\ref{eq:2}) is then simply
\begin{equation}
f(X)=f_+(X^+)+f_-(X^-)\ .
\end{equation}
The energy-momentum tensor (\ref{eq:energy-momentum}) takes the form
\begin{equation}
T_{\pm\pm}(X^\pm)=\left(f_{,\pm}(X^\pm)\right)^2\ ,\quad T_{+-}=0\,.
\label{eq:nullt}
\end{equation}
and (\ref{eq:3}) can be solved for $y$:
\begin{equation}
y(X)=\kappa^2 X^+ X^- + y_+(X^+) + y_-(X^-)\ .
\label{eq:y1}
\end{equation}
Here
\begin{equation}
y_\pm(X^\pm)=-\int^{X^\pm} d\bar X{}^\pm\int^{\bar X{}^\pm}
d\bar{\bar X}^\pm\ T_{\pm\pm}(\bar{\bar X}{}^\pm)\ .
\label{eq:y2}
\end{equation}
Note that the only non-trivial dynamics is contained in the scalar field 
$f$.
The spacetime metric $\gamma_{\alpha\beta}$ is flat, while 
(\ref{eq:y1})-(\ref{eq:y2}) show that the dilaton field $y$ is completely
determined (up to constants of integration) by $f$.
The physical metric (\ref{eq:physical_metric}) is dynamical via its 
dependence on $y$.
Singularities in $\bar\gamma_{\alpha\beta}$ usually occur where $y=0$.

\subsection*{2b. Canonical form of the dilatonic action}

The spacetime action (\ref{eq:S}) is cast into canonical form by 
the standard ADM decomposition \cite{adm}.
Given an arbitrary foliation $X^\alpha=X^\alpha(t,x^a)$ of a spacetime
by ($t={\rm const}$) spacelike hypersurfaces, one has the general
decomposition formula \cite{KKjmp}:
\begin{equation}
|\gamma|^{{1\over 2}}R[\gamma]=N g^{{1\over 2}}
\left(K_{ab}K^{ab}-K^2+R[g]\right)-2 g^{{1\over 2}}\triangle_g N + 
2(g^{{1\over 2}}KN^a)_{,a}-2(g^{{1\over 2}}K)\dot{}\ .
\end{equation}
In 1+1 dimensions, this reduces to 
\begin{equation}
|\gamma|^{{1\over 2}}R[\gamma]=
-2 g^{{1\over 2}}\triangle_g N + 2(g^{{1\over 2}}KN^1)'
-2(g^{{1\over 2}}K)\dot{}\ ,
\label{eq:decomp}
\end{equation}
where the prime $'$ denotes partial derivative with respect to the 
(single) spatial coordinate $x \; ( x\in (-\infty , \infty))$, and 
the dot $\dot{}$ denotes partial 
derivative with respect to the time coordinate $t$.
Here, $g$ is the determinant of the induced spatial metric, and $K$ is the
trace of the extrinsic curvature of the $t={\rm const}$ hypersurfaces.
$N$ and $N^1$ are the lapse function and shift vector, respectively.

Since the induced spatial metric is 1-dimensional, it has only one
independent component
\begin{equation}
\sigma^2:=g_{11}=X^\alpha{}'X^\beta{}'\gamma_{\alpha\beta}\ .
\end{equation}
Similarly, the extrinsic curvature tensor is completely specified by 
\begin{equation}
K_{11}=-X^\alpha{}'X^\beta{}'n_{\alpha;\beta}\ ,
\end{equation}
where $n_\alpha$ is the unit (covariant) timelike normal to the
spacelike hypersurfaces.
Since in one dimension tensor fields of contravariant rank $r$ and 
covariant rank $s$ transform as scalar densities of weight $(s-r)$, 
%$\scriptstyle{\left(\begin{array}{c}r\\s\end{array}\right)}$
it follows that $\sigma$ transforms as a scalar density of weight
$+1$, while $K:=g^{11}K_{11}$ and the lapse function $N$ transform 
as ordinary scalars.
The shift vector $N^1$ transforms as a scalar density of weight $-1$.

If we substitute (\ref{eq:decomp}) into the spacetime action (\ref{eq:S}),
integrate by parts, and discard the boundary terms, we find
\begin{eqnarray}
\lefteqn{S[y,\sigma,f,N,N^1]}\nonumber\\
&&=\int dt\int dx\ \left(\sigma K(\dot y-N^1 y')
-N(\sigma\triangle_g y-2\kappa^2\sigma)\right)+{\rm matter}\ ,
\label{eq:Slagrangian}
\end{eqnarray}
where $\sigma K$ may be thought of as shorthand notation for
\begin{equation}
\sigma K= N^{-1}\left(-\dot\sigma+(N^1\sigma)'\right)\ .
\end{equation}
The matter contribution to the action is given by
\begin{equation}
{\rm matter}=\int dt\int dx\ \left(-{1\over 2}N\sigma^{-1}f'{}^2
+{1\over 2}N^{-1}\sigma(\dot f-N^1 f')^2\right)\ .
\label{eq:matter}
\end{equation}
As usual, $N$ and $N^1$ play the role of Lagrange multipliers of the 
theory.
The dynamical variables are $y$, $\sigma$, and $f$.

The momenta conjugate to $y$, $\sigma$, and $f$ are
\begin{eqnarray}
\pi_y&=&\sigma K=N^{-1}\left(-\dot\sigma+(N^1\sigma)'\right)\ ,\\
p_\sigma&=&N^{-1}(-\dot y+ N^1 y')\ ,\label{eq:psigmay}\\
\pi_f&=&N^{-1}\sigma(\dot f - N^1 f')\ .
\end{eqnarray}
The notation we have chosen is such that the canonical variables denoted 
by Latin symbols $(y,f,p_\sigma)$ transform as ordinary scalars, 
while those denoted by Greek symbols $(\sigma,\pi_y,\pi_f)$ 
transform as scalar densities of weight $+1$.
The above equations for the momenta can be inverted, yielding expressions 
for the velocities in terms of the momenta.
It is then a straightforward exercise to cast the action
(\ref{eq:Slagrangian})-(\ref{eq:matter}) into Hamiltonian form
\begin{eqnarray}
\lefteqn{S[y,\pi_y,\sigma,p_\sigma,f,\pi_f,N,N^1]}\nonumber\\
&&=\int dt\int dx\ \left(\pi_y\dot y+p_\sigma\dot\sigma+
\pi_f\dot f-N H - N^1 H_1\right)\ ,
\label{eq:Shamiltonian}
\end{eqnarray}
where
\begin{eqnarray}
H&:=&-\pi_y p_\sigma + \sigma\triangle_g y -2 \kappa^2\sigma
+{1\over 2}\sigma^{-1}(\pi_f^2+f'{}^2)\ ,\\
H_1&:=&\pi_y y' - \sigma p_\sigma' + \pi_f f'\ 
\end{eqnarray}
are the super-Hamiltonian and supermomentum, which are constrained to
vanish $(H\approx 0 \approx H_1)$ as a consequence of variations of 
$N$ and $N^1$.

In what follows, it is more convenient to work with a rescaled
super-Hamiltonian and lapse function:
\begin{equation}
\bar H:=\sigma H\ ,\quad\bar N:=\sigma^{-1}N\ .
\end{equation}
Since
\begin{equation}
\sigma^2\triangle_g y:=\sigma^2 g^{-{1\over 2}}
(g^{{1\over 2}}g^{11} y')'=y''-\sigma^{-1}\sigma'y'\ ,
\end{equation}
we have
\begin{equation}
\bar H=-\pi_y\sigma p_\sigma + y''-\sigma^{-1}\sigma'y'
- 2 \kappa^2\sigma^2 +{1\over 2}(\pi_f^2+f'{}^2)\ .
\end{equation}
Both $\bar H$ and $H_1$ transform as scalar densities of weight $+2$.

\subsection*{2c. Canonical transformation to embedding variables}

As shown in section 2a, the equations of motion imply that the 
spacetime metric $\gamma_{\alpha\beta}$ is flat:
\begin{equation}
ds^2=dX^+ dX^-\ .
\end{equation}
This means that given an arbitrary foliation $X^\pm=X^\pm(t,x)$ of 
spacetime by ($t={\rm const}$) spacelike hypersurfaces, the induced 
spatial metric and the trace of the extrinsic curvature are
\begin{eqnarray}
g_{11}&=&X^+{}'X^-{}'\ ,
\label{eq:g11}\\
g^{{1\over 2}}K&=&-{1\over 2}
\left[\ln\left({X^+{}'\over X^-{}'}\right)\right]^{'} \ .
\label{eq:gK}
\end{eqnarray}
Since $\sigma^2=g_{11}$ and $\pi_y=\sigma K$, equations 
(\ref{eq:g11}) and (\ref{eq:gK}) imply
\begin{eqnarray}
\sigma&=& \sqrt{X^+{}'X^-{}'}\, , \label{eq:sigmaX}\\
\pi_y&=&-{1\over 2}\left[\ln\left({X^+{}'\over X^-{}'}\right)\right]^{'} \ .
\label{eq:pi_yX}
\end{eqnarray}

The connection (\ref{eq:sigmaX})-(\ref{eq:pi_yX}) between the null 
coordinates $X^\pm$ and the geometry of embeddings $\sigma$, $\pi_y$ holds 
only modulo the field equations.
This does not prevent us, however, from introducing the embedding variables 
$X^\pm(x)$ as new canonical coordinates on phase space such that 
(\ref{eq:sigmaX})-(\ref{eq:pi_yX}) are satisfied even prior to varying
the action.
Following this strategy, we want to complete 
(\ref{eq:sigmaX})-(\ref{eq:pi_yX}) into a canonical transformation
\begin{equation}
\Big(y(x),\pi_y(x),\sigma(x),p_\sigma(x)\Big)\leftrightarrow 
\Big(X^\pm(x),\Pi_\pm(x)\Big)\ .
\end{equation}
Since the matter variables remain unchanged, they are not mentioned in 
what follows.
In this section, we will ignore important issues related to falloff
conditions on the field variables, and corresponding boundary terms
and constants of integration.
These issues will be addressed in section 3, where we give a detailed
discussion of asymptotics.

We proceed in a series of steps:

\begin{itemize}
\item[(i)] 
We first replace $(y,\pi_y)$ by $(z,\pi_z)$ via the canonical
transformation
\begin{eqnarray}
y(x)&=& -\int^x d\bar x\ z(\bar x)\pi_z(\bar x)\ , \label{eq:37}\\
\pi_y &=& -(\ln z)'\ . \label{eq:38}
\end{eqnarray}
This brings the super-Hamiltonian $\bar H$ into a more symmetric form
\begin{equation}
\bar H=z^{-1}z'\sigma p_\sigma + \sigma^{-1}\sigma' z\pi_z-(z\pi_z)' 
- 2\kappa^2\sigma^2+{1\over 2}(\pi_f^2+f'{}^2)\ .
\end{equation}
\item[(ii)] 
We then express $(\sigma,z)$ as combinations of the density 
variables $\xi^\pm$:
\begin{eqnarray}
\sigma&=& \sqrt{\xi^+\xi^-}\ ,
\label{eq:sigma1}\\
z&=& \sqrt{{\xi^+\over \xi^-}}\ .
\label{eq:z}
\end{eqnarray}
Equations (\ref{eq:sigma1}), (\ref{eq:z}) can be completed into a 
point canonical transformation 
$(\sigma,p_\sigma,z,\pi_z)\leftrightarrow(\xi^\pm,p_\pm)$, 
with
\begin{eqnarray}
\sigma p_\sigma &=& \xi^+ p_+ +\xi^- p_-\ ,\\
z\pi_z &=& \xi^+ p_+ -\xi^- p_-\ . \label{eq:43}
\end{eqnarray}
Because
\begin{eqnarray}
{\sigma'\over\sigma} &=& {1\over 2}\left({\xi^+{}'\over\xi^+} + 
{\xi^-{}'\over\xi^-}\right)\ ,\\
{z'\over z} &=& {1\over 2}\left({\xi^+{}'\over\xi^+} -
{\xi^-{}'\over\xi^-}\right)\ ,
\end{eqnarray}
$\bar H$ assumes the form
\begin{equation}
\bar H=-\xi^+ p_+{}' +\xi^- p_-{}'-2\kappa^2 \xi^+\xi^-
+{1\over 2}(\pi_f^2+f'{}^2)\ .
\end{equation}
\item[(iii)]
We interchange the roles of the coordinates $\xi^\pm$ and momenta 
$p_\pm$ by the canonical transformation 
$(\xi^\pm,p_\pm)\leftrightarrow(X^\pm,\bar\Pi_\pm)$:
\begin{eqnarray}
\xi^\pm &=& X^\pm{}'\ , \label{eq:47}\\
p_\pm(x) &=& -\int^x d\bar x\ \bar\Pi_\pm(\bar x)\ . \label{eq:48}
\end{eqnarray}
This gives
\begin{equation}
\bar H=\bar\Pi_+ X^+{}'-\bar\Pi_- X^-{}'
-2\kappa^2 X^+{}'X^-{}'+{1\over 2}(\pi_f^2+f'{}^2)\ .
\end{equation}
\item[(iv)] 
Finally, we absorb the term $2\kappa^2 X^+{}'X^-{}'$ by a redefinition 
of the momenta $\bar\Pi_\pm\rightarrow\Pi_\pm$:
\begin{equation}
\bar\Pi_\pm=\Pi_\pm \pm{\kappa}^{2} X^\mp{}'\ . \label{eq:50}
\end{equation}
This yields
\begin{equation}
\bar H=\Pi_+ X^+{}' -\Pi_- X^-{}'+{1\over 2}(\pi_f^2+f'{}^2)\ .
\label{eq:Hbar}
\end{equation}
Since $X^\pm$ and $f$ transform as ordinary spatial scalars, the 
supermomentum $H_1$ necessarily takes the form
\begin{equation}
H_1=\Pi_+X^+{}'+\Pi_- X^-{}'+\pi_f f'\ .
\label{eq:H1}
\end{equation}
\end{itemize}

\noindent
The super-Hamiltonian and supermomentum constraints can then be 
combined into the Virasoro pair:
\begin{equation}
H^\pm:={1\over 2}(\bar H\pm H_1)
=\pm\Pi_\pm X^\pm{}' +{1\over 4}(\pi_f\pm f')^2\approx 0\ .
\label{eq:vc}
\end{equation}
Our derivation of the canonical transformation 
(\ref{eq:37})-(\ref{eq:38}), (\ref{eq:sigma1})-(\ref{eq:43}),
(\ref{eq:47})-(\ref{eq:48}), and (\ref{eq:50}) makes it clear that 
the embedding variables $X^{\pm}(x)$ on the space of solutions really 
describe embeddings in the double null Minkowskian coordinates 
$X^{\pm}$ of the flat background metric $\gamma_{\alpha \beta}$.

\subsection*{2d. Commuting constraints and Heisenberg variables}

By scaling (\ref{eq:vc}), we obtain an equivalent set of commuting
constraints:
\begin{equation}
{\bf\Pi}_\pm(x):=\Pi_\pm(x)+h_\pm(x;X^\pm,f,\pi_f]\approx 0\ ,
\label{eq:comcon}
\end{equation}
where
\begin{equation}
h_\pm(x;X^\pm,f,\pi_f]:=\pm{1\over 4}\left(X^\pm{}'(x)\right)^{-1}
\left(\pi_f(x)\pm f'(x)\right)^2\ ,
\label{eq:eflux}
\end{equation}
and
\begin{equation}
\{{\bf\Pi}_\pm(x),{\bf\Pi}_\pm(\bar x)\}=0=
\{{\bf\Pi}_+(x),{\bf\Pi}_-(\bar x)\}\ .
\label{eq:com}
\end{equation}
Because
\begin{equation}
X^\pm{}'(x)f_{,\pm}(X^\pm(x))={1\over 2}\left(f'(x)\pm\pi_f(x)\right)\ ,
\label{eq:fcan}
\end{equation}
the energy flux (\ref{eq:eflux}) is simply related to the null components
(\ref{eq:nullt}) of the energy-momentum tensor:
\begin{equation}
h_\pm(x;X^\pm,f,\pi_f]=\pm X^\pm{}'(x)T_{\pm\pm}(X^\pm(x))\ .
\label{eq:h-T}
\end{equation}

Because the dynamical variables ${\bf\Pi}_\pm(x)$ commute, (\ref{eq:com}), 
we can choose them as new momenta.
The embedding variables remain their canonically conjugate coordinates:
\begin{equation}
{\bf X}^\pm(x)=X^\pm(x)\ .
\label{eq:h=sx}
\end{equation}
To complete the canonical variables $({\bf X}^\pm(x),{\bf\Pi}_\pm(x))$
into a canonical chart on phase space, we need to replace the conjugate 
field variables $(f(x),\pi_f(x))$ by new conjugate variables
$({\bf f}(x),{\bf\pi}_{{\bf f}}(x))$ which commute with 
${\bf X}^\pm(x)$ and ${\bf\Pi}_\pm(x)$.
The commutation with ${\bf\Pi}_\pm(x)$, i.e., with the constraints, means 
that ${\bf f}(x)$ and ${\bf\pi}_{{\bf f}}(x)$ are constants of the motion.
They can be identified with initial data on a fixed embedding
$X^\pm(x)=X^\pm_{(0)}(x)$.
In flat spacetime, it is natural to choose $X^\pm_{(0)}(x)$ as the
$T=0$ hypersurface parametrized by the Cartesian coordinate $Z$:
\begin{equation}
X^\pm_{(0)}(x)=x\ .
\label{eq:inem}
\end{equation}
The canonical transformation
\begin{equation}
\left(X^\pm(x),\Pi_\pm(x),f(x),\pi_f(x)\right)\leftrightarrow
\left({\bf X}^\pm(x),{\bf\Pi}_\pm(x),{\bf f}(x),{\bf \pi}_{\bf f}(x)
\right)
\label{eq:he-sch}
\end{equation}
is closely connected with the passage from the Schr\"odinger to the
Heisenberg picture.
We shall call the boldface canonical variables the fundamental Heisenberg
variables, and the lightface canonical variables the fundamental
Schr\"odinger variables.
The canonical transformation between the respective embedding coordinates
and momenta is given by (\ref{eq:h=sx}) and (\ref{eq:comcon}).
When we write the commutation relation
$\{f(x),\Pi_\pm(\bar x)\}=0$ between the Schr\"odinger variables in the
Heisenberg canonical chart, we get the Heisenberg equation of 
motion
\begin{equation}
{\delta f(x;{\bf X},{\bf f},{\bf\pi}_{{\bf f}}]\over
\delta{\bf X}^\pm(\bar x)}=
\{f(x;{\bf X},{\bf f},{\bf\pi}_{{\bf f}}],
h_\pm(\bar x;{\bf X}^\pm,{\bf f},{\bf\pi}_{{\bf f}}]\}
\label{eq:heq}
\end{equation}
for the field $f(x;{\bf X},{\bf f},{\bf\pi}_{{\bf f}}]$.
A similar equation holds for the field momentum
$\pi_f(x;{\bf X},{\bf f},{\bf\pi}_{{\bf f}}]$.
The solution of the Heisenberg equations of motion under the initial
condition that $f(x)$, $\pi_f(x)$ match ${\bf f}(x)$, ${\bf\pi}_{{\bf f}}(x)$ 
at the initial embedding $X^\pm(x)=X^\pm_{(0)}(x)$ gives the canonical 
transformation from the Heisenberg to the Schr\"odinger field data.

\section*{3. Taking care of asymptotic conditions}

In the transformation to the embedding variables $(X^\pm(x),\Pi_\pm(x))$,
we did not pay any attention to boundary terms at the two spatial infinities. 
In this section we shall complete the analysis by evaluating these 
contributions. 
To do this, we need to know the asymptotic behavior of the phase space 
variables (section 3a).
These are easiest to motivate for the embedding variables 
$(X^\pm(x),\Pi_\pm(x))$.
We then choose the falloff conditions on the multipliers 
$({\bar N}(x),N^1(x))$ to ensure differentiability of the Hamiltonian.
The falloff conditions for the original geometric variables 
$(y(x),\pi_y(x),\sigma(x),p_{\sigma}(x))$ are deduced from the 
asymptotic behavior of the spacetime solution.
In section 3b, we show that the transformation between the geometric 
and embedding variables that respects their falloff conditions 
requires parametrization of the infinities by proper times $\tau_L$,
$\tau_R$, and complementation of the embedding variables by a new canonical 
pair $m_R$, $p$:
\begin{equation}
\Big(y(x),\pi_y(x),\sigma(x),p_{\sigma}(x);\tau_L,\tau_R\Big)
\leftrightarrow\Big(X^\pm(x),\Pi_\pm(x);m_R,p\Big)\ .
\label{eq:ct}
\end{equation}
The physical meaning of all of these variables will naturally emerge from
the examination of the action.

In the rest of the paper, we distinguish the terms at the left and right
infinities by the subscripts $L$ and $R$.
Their suppression means that the discussion is valid at both infinities.

\subsection*{3a. Asymptotic conditions}

Before discussing the embedding  variables, let us briefly discuss the 
asymptotic behavior of the matter variables. 
With a view towards quantum theory, we impose such falloff conditions 
that the Klein-Gordon symplectic norm associated with points on the matter 
phase space is well-defined. 
To achieve this, it suffices to require that $f(x),\pi_f(x)\in{\cal S}$
lie in the Schwartz space ${\cal S}$ of smooth functions with rapid
decay at infinity.
One can check that with these conditions on the matter variables
and our subsequent choice of falloff conditions on the multipliers 
${\bar N}(x)$, $N^1(x)$ the matter part of the action is well-defined and 
differentiable on phase space.

Falloff conditions on the embedding variables $X^\pm(x)$ are chosen by 
examining the asymptotic behavior of the classical solution 
(\ref{eq:y1})-(\ref{eq:y2}) for $y$. 
To leading order in $X^\pm(x)$ at the spatial infinities
\begin{equation}
y(x)=\left(\kappa X^+(x) +{A^+\over\kappa}\right)
\left( \kappa X^-(x) + {A^-\over\kappa}\right)  
+ {m\over \kappa}\ ,
\label{eq:ylarge}
\end{equation}
where $A^\pm$ and $ m $ are constants. 
Equation (\ref{eq:ylarge}) can be verified by substituting the falloff 
conditions of the scalar field variables $f(x)$, $\pi_f(x)$ into the explicit 
solution for $y(x)$.
The form of (\ref{eq:ylarge}) (although not the values of $A^\pm  $)
is left invariant by translations 
\begin{equation}
X^{\pm}(x) \rightarrow X^{\pm}(x) + \xi^{\pm}
\label{eq:tran}
\end{equation}
of the Minkowskian coordinates $X^\pm(x)$, and by their boosts
\begin{equation}
X^\pm(x) \rightarrow e^{\pm \tau} X^\pm(x)\ .
\label{eq:atts}
\end{equation}
The physical interpretation of these transformations is as follows:
The physical metric, $\bar\gamma_{\alpha\beta}$, is asymptotically flat.
A clock moving near spatial infinity along the orbits of the asymptotic
stationary Killing field of the physical metric measures the parameter 
$\tau$.
The transformations (\ref{eq:atts}) correspond to asymptotic time
translations along the orbits of this Killing field.
We will refer to $\tau$ as the Killing time measured at infinity.
We fix the  {\it Minkowskian} translational 
freedom (\ref{eq:tran}) by imposing $A^{+}_{R} = 0 =A^{-}_L$.

A deeper analysis of the role of $A^\pm $ in the canonical theory reveals 
that they are related to the generators of translations (\ref{eq:tran})
 of $X^\pm(x)$
(see (\ref{eq:APi1})-(\ref{eq:APi2})).
Hence, in the canonical description, setting the values of 
$A^{+}_{R}$ and $A^{-}_L $ to zero must be accompanied by freezing the 
Minkowskian
translational freedom in $X^+(x)$ at right infinity and $X^-(x)$ at left 
infinity. 
Thus we require that
\begin{eqnarray}
X^{-}(x) & = & e^{-\tau{_L}} x + O(x^{-2})\ ,
\label{eq:Xleft}\\
X^{+}(x) & = & e^{\tau{_L}} x + \xi^{+}_L + O(x^{-2})
\end{eqnarray}
near left infinity, and
\begin{eqnarray}
X^{-}(x) & = & e^{-\tau} x + \xi^{-}_R + O(x^{-2})\ ,\\ 
X^{+}(x) &  =  & e^{\tau} x + O(x^{-2})  
\label{eq:Xright}
\end{eqnarray}
near right infinity.
Here, $\tau_L$ and  $\tau$ are the Killing times of the physical geometry
at the left and right infinities.\footnote{The symbol 
$\tau$ is an exception to the $L$ and $R$ subscript convention; 
i.e., even though $\tau$ is the Killing time at $+\infty$, it carries no 
$R$ subscript. 
The symbol $\tau_R$ is reserved for the parametrization time at $+\infty$, 
which will be introduced later.

We can make the parametrization time and the Killing time coincide at one 
of the infinities, but not at both. 
We shall later choose to make them coincide at the left infinity. 
Hence $\tau_L$ has a dual interpretation of Killing time and 
parametrization time at $-\infty$, while $\tau$ and $\tau_R$ are 
physically distinct quantities associated with $+\infty$.}
The parameters $\xi^{+}_{L}$ and $\xi^{-}_{R}$ correspond to the residual
translational freedom (\ref{eq:tran}) 
in $X^+(x)$ at left infinity and $X^-(x)$ at right 
infinity. 

For $\Pi_\pm(x)$ to generate cotangent maps from the space of $X^\pm(x)$,
we require that
\begin{equation}
\Pi_{\pm}(x) = O(x^{-3})
\label{eq:Pi}
\end{equation}
at both infinities.

For the smeared constraint functionals 
\begin{equation}
\bar H[\bar N]:=\int_{-\infty}^\infty dx\ \bar N(x)\bar H(x)\ ,
\quad
H_1[N^1]:=\int_{-\infty}^\infty dx\ N^1(x) H_1(x)
\label{eq:smconstr}
\end{equation}
to be functionally differentiable and to preserve 
(\ref{eq:Xleft})-(\ref{eq:Pi}), we put
\begin{eqnarray}
\bar N(x)&=&  \alpha_L x +\nu_L + O(x^{-2})\ ,
\label{eq:Nleft}\\
N^1(x)&=& \nu_L + O(x^{-2})  
\end{eqnarray}
at the left infinity, and
\begin{eqnarray}
\bar N(x)&=& \alpha_R x - \nu_R + O(x^{-2})\ ,\\
N^1(x)&=& \nu_R + O(x^{-2})
\label{eq:Nright}
\end{eqnarray}
at the right infinity.
Note that the infinitesimal changes (generated by the constraint functionals) 
of $(\tau_L,\tau,\xi^+_L,\xi^-_R)$ are $(\alpha_L,\alpha_R,\nu_L,\nu_R)$.

Next, we  motivate and state our choice of  the asymptotic behavior of 
the geometric variables $(y(x),\pi_y(x),\sigma(x),p_\sigma(x))$.
By substituting the falloff conditions (\ref{eq:Xleft})-(\ref{eq:Xright}) 
in the asymptotic form  of the spacetime solution (\ref{eq:ylarge}) for 
$y(x)$,  we obtain
\begin{equation}
y(x) =  {\kappa}^2 x^2  + B_L  x
+ {m_L\over \kappa}+ O(x^{-1}) \label{eq:yleft}
\end{equation}
at the left infinity, and 
\begin{equation}
y(x) =  {\kappa}^2 x^2  + B_R x 
+ {m_R\over \kappa} + O(x^{-1}) \label{eq:yright}
\end{equation}
at the right infinity.
Here 
\begin{eqnarray}
B_L&=&(A^+_{L} + \kappa^2 \xi^+_L) e^{-\tau_L}\ , \\
B_R&=&(A^-_{R} + \kappa^2 \xi^-_R) e^{\tau}\ , 
\end{eqnarray}
and $m_L$ and $m_R$ are the values of the parameter $m$ in (\ref{eq:ylarge}) 
at the left and right infinities.

By substituting the falloff conditions
(\ref{eq:Xleft})-(\ref{eq:Xright}) in (\ref{eq:pi_yX}) and 
(\ref{eq:sigmaX}) for $\pi_y(x)$ and $\sigma(x)$, we obtain
\begin{eqnarray}
\pi_y(x)& = & O(x^{-4})\ ,
\label{eq:piyLR}\\
\sigma(x)& = & 1 + O(x^{-3}) 
\label{eq:sigmaLR}
\end{eqnarray}
at both infinities.

Finally, by substituting the falloffs (\ref{eq:yleft}), (\ref{eq:yright}),
and (\ref{eq:Nleft})-(\ref{eq:Nright}) in (\ref{eq:psigmay}), we obtain
\begin{equation}
p_{\sigma}(x) =  B_L + O(x^{-2}) 
\label{eq:psigmaL}
\end{equation}
at the left infinity, and 
\begin{equation}
p_{\sigma}(x) =  -B_R + O(x^{-2}) 
\label{eq:psigmaR}
\end{equation}
at the right infinity.

Equations (\ref{eq:yleft})-(\ref{eq:psigmaR}) constitute our choice of 
boundary conditions on the $(y(x),\pi_y(x),\sigma(x),p_\sigma(x))$ variables. 
Although they have been deduced from the asymptotic behavior of a 
{\it solution} to the field equations, it can be checked that their 
imposition on the {\it entire} phase space leads to a consistent 
Hamiltonian formulation. 
In particular, all the relevant Hamiltonian flows preserve these falloff 
conditions.

\subsection*{3b. Canonical action in the geometric and embedding variables}

It can be checked that with the asymptotic conditions
(\ref{eq:yleft})-(\ref{eq:psigmaR}) on the geometric variables, and the 
falloff conditions
(\ref{eq:Nleft})-(\ref{eq:Nright}) on the lapse and shift multipliers, 
the  action  (\ref{eq:Shamiltonian}) must be complemented by a surface 
term to be functionally differentiable: 
\begin{eqnarray}
\lefteqn{S[y,\pi_y,\sigma,p_\sigma,f,\pi_f,{\bar N},N^1]}\nonumber\\
&=&\int dt\int_{-\infty}^\infty dx\ \left(\pi_y\dot y+p_\sigma\dot\sigma
+\pi_f\dot f-{\bar N}{\bar H} - N^1H_1\right)\nonumber\\
&&\quad\quad\quad +\int dt\ \left(-\alpha_R{m_R\over\kappa}
+\alpha_L{m_L\over \kappa}\right)\ .
\label{eq:Sbdry}
\end{eqnarray}
We see that the mass parameters generate the asymptotic Killing time 
translations (\ref{eq:atts}) of the physical metric at the left and right
infinities.
{}From the viewpoint of canonical Hamiltonian theory, this property of $m_L$ 
and $m_R$ justifies their identification with the left and right mass of the
system.\footnote{Our identification of the generators of the transformations
(\ref{eq:atts}) with the mass of the system differs from that made by
Mikovi\'{c} \cite{mik 3}.
In \cite{mik 3}, certain combinations of the generators of 
Minkowskian translations (\ref{eq:tran})
of $X^\pm(x)$ are identified with the mass.}
The equations of motion which follow from this action preserve our choice 
of asymptotic behavior of the $(y(x),\pi_y(x),\sigma(x),p_\sigma(x))$ 
variables. 

A `boundary action' similar to that in (\ref{eq:Sbdry}) 
appears  in the study of the Schwarzschild black holes \cite{KKSchw}.
By following the method introduced in \cite{KKSchw,MV}, we parametrize the 
asymptotic time translations (\ref{eq:atts})
at the spatial infinities by introducing into 
the action two additional parameters $\tau_R$ and $\tau_L$:
\begin{eqnarray}
\lefteqn{S[y,\pi_y,\sigma,p_\sigma,f,\pi_f,{\bar N},N^1;\tau_L,\tau_R)}
\nonumber\\
&=&\int dt\int_{-\infty}^\infty dx\ \left(\pi_y\dot y+p_\sigma\dot\sigma 
+\pi_f\dot f-{\bar N}{\bar H} - N^1 H_1\right)\nonumber \\
&&\quad\quad\quad+\int dt\ \left( -{\dot \tau_R}{m_R\over \kappa} 
+{\dot \tau_L} {m_L\over \kappa}\right)\,.
\label{eq:Sparam}
\end{eqnarray}
Note that the reading $\tau_R$ of the right parametrization clock
does not necessarily coincide with the proper time $\tau$ identified
from the geometry.
The equations of motion, however, insure that $\tau_R$ and $\tau$ are
running at the same rate, so that the difference is due only to an initial
setting.

The dynamical variables in the action (\ref{eq:Sparam}) are the 
original field variables $(y(x),\pi_y(x),\sigma(x),p_\sigma(x))$ and the  
parameters $(\tau_L,\tau_R)$. 
To define a one-to-one, invertible transformation to the embedding 
variables, we need to complement $(X^\pm(x),\Pi_\pm(x))$  by a 
pair of parameters. 
One of these is simply the mass parameter $m_R$ at the right infinity. 
The second parameter is 
\begin{equation}
p:= {{\tau_R - \tau} \over \kappa}\ .
\label{eq:pdef}
\end{equation}
Its physical meaning will follow from the transformation equations.
We can cast the parametrized action (\ref{eq:Sparam}) into manifestly 
canonical form by transforming 
$\left(y(x),\pi_y(x),\sigma(x),p_\sigma(x);\tau_L,\tau_R\right)$ into
$\left(X^\pm(x),\Pi_\pm(x);m_R,p\right)$:
\begin{eqnarray}
y(x)&=&{\kappa}^2 X^+(x)X^-(x) 
\nonumber\\
&&-\int^{x}_{\infty}d\bar x\ X^-{}'(\bar x)
\int^{\bar x}_{\infty}d\bar{\bar x}\ \Pi_-(\bar{\bar x})
+\int^{x}_{-\infty}d\bar x\ X^+{}'(\bar x)
\int^{\bar x}_{-\infty}d\bar{\bar x}\ \Pi_+(\bar{\bar x})
\nonumber\\
&&+\int^{\infty}_{-\infty}dx\ X^+(x)\Pi_+(x) +{m_R\over\kappa}\ ,
\label{eq:y(X)}\\
\pi_y(x)&=&-{1\over 2}\left[\ln\left({X^+{}'(x)\over X^-{}'(x)}\right)
\right]^{'}\ ,\\
\sigma(x)&=&\sqrt{X^+{}'(x)X^-{}'(x)}\ ,\\
p_{\sigma}(x)&=&{1\over \sqrt{X^+{}'(x)X^-{}'(x)}} 
\Bigg({\kappa}^2\bigg(X^+(x)X^-{}'(x)- X^+{}'(x)X^-(x)\bigg)\nonumber\\
&&-X^-{}'(x)\int^{x}_{\infty}d\bar x\ \Pi_-(\bar x) - 
X^+{}'(x)\int^{x}_{-\infty}d\bar x\ \Pi_+(\bar x)\Bigg)\ ,\\
\tau_L&=&-\lim_{x\rightarrow -\infty}\ln\left({{X^- (x)}\over x}\right)\ ,\\
\tau_R&=&\kappa\,p+\lim_{x\rightarrow\infty}\ln\left({{X^+ (x)}\over x}
\right)\ .
\label{eq:tau(X)}
\end{eqnarray}
(The above transformation equations are obtained by following steps 
(i)-(iv) in section 2c, paying proper attention to boundary terms.)
It can be checked (using (\ref{eq:Xleft})-(\ref{eq:Pi})
and (\ref{eq:y(X)})-(\ref{eq:tau(X)})) that up to finite total time 
derivatives
\begin{eqnarray}
\lefteqn{\int_{-\infty}^{\infty}dx\ (\pi_y{\dot y} + p_\sigma\dot\sigma)
- {\dot \tau_R}{m_R\over \kappa}+{\dot \tau_L}{m_L\over \kappa}}\nonumber\\
&=&\int_{-\infty}^\infty dx\ (\Pi_+\dot X{}^+ + \Pi_-\dot X{}^-)
+ p\dot m_R\ .
\label{eq:liouform}
\end{eqnarray}
It is best to confine $x$ to a finite interval $x_L<x<x_R$ and to take
the limits $x_L \rightarrow -\infty$ and $x_R \rightarrow \infty$ only at
the end.

To summarize, from equations  (\ref{eq:Sparam}) and (\ref{eq:liouform})
we conclude that up to an unimportant finite total time derivative 
\begin{eqnarray}
\lefteqn{S[y,\pi_y,\sigma,p_\sigma,f,\pi_f,{\bar N},N^1;\tau_L,\tau_R)}
\nonumber\\
&=&S[X^\pm,\Pi_\pm,f,\pi_f,{\bar N},N^1;p,m_R)\nonumber\\
&=&\int dt\int_{-\infty}^{\infty}dx\ \left(\Pi_+\dot X{}^+ +
\Pi_-\dot X{}^-+\pi_f\dot f-{\bar N}{\bar H} - N^1 H_1\right)\nonumber\\
&&\quad\quad\quad +\int dt\ p\dot m_R\ ,
\label{eq:Sfinal}
\end{eqnarray}
with ${\bar H}$, $H_1$ taking the form of the constraints for a 
parametrized massless scalar field on a 2-dimensional Minkowski spacetime 
as in (\ref{eq:Hbar}) and (\ref{eq:H1}). 

It is easy to see that the right mass $m_R$ and its conjugate momentum 
$p$ are constants of the motion. 
{}From (\ref{eq:pdef}), we see that $p$ can be interpreted as the difference 
between the  proper time $\tau_R$ as measured by the right parametrization 
clock and the proper time $\tau$ reconstructed from the 
geometry.\footnote{For more about this interpretation see \cite{KKSchw,MV}.
In fact, the parameters $\tau$, $\tau_L$, $\tau_R$ correspond to those 
referred to as Killing time and left and right parametrization time in 
\cite{MV}.}

Finally, we can express the parameters $A^+_L$, $A^-_R$ and the left mass
$m_L$ in terms of the new set of variables:
\begin{eqnarray}
A_L^+&=&\int _{-\infty}^\infty dx\ \Pi_-(x) \ ,
\label{eq:APi1}\\
A_R^-&=&\int _{-\infty}^\infty dx\ \Pi_+(x)\ , 
\label{eq:APi2}
\end{eqnarray}
and
\begin{equation}
{m_L\over\kappa}={m_R\over\kappa}
+\int_{-\infty}^\infty dx\ X^+(x)\Pi_+(x)
-\int_{-\infty}^\infty dx\ X^-(x)\Pi_-(x)\ .
\label{eq:mL}
\end{equation}
Equations (\ref{eq:APi1})-(\ref{eq:APi2}) show that $A_L^+$ and $A_R^-$ 
 generate Minkowskian translations (\ref{eq:tran}) of $X^-(x)$ and $X^+(x)$, 
respectively, through the Poisson brackets.

\section*{4. Constraint quantization of the dilatonic model}

The constraints (\ref{eq:vc}) of dilatonic gravity
have the same form as those of a parametrized massless scalar field 
propagating on a flat 2d background.
This reflects the conformal invariance of the scalar wave equation in two 
dimensions:
While the scalar field curves the spacetime in which it propagates, the 
curvature does not affect the propagation.
Because every 2-geometry is conformally flat, one can consider the
propagation as taking place on an auxiliary flat background, rather than 
in the physical curved spacetime.
The way in which we have reconstructed the embedding variables $X^\pm(x)$ 
from the geometric variables $g_{11}$ and $K_{11}$ guarantees that they 
coincide with the (double-null) Cartesian coordinates on this auxiliary 
background.\footnote{The variables $X^\pm(x)$ used by Cangemi et 
al.~\cite{jac 3} do not all 
have the meaning of embedding variables, but they are 
rather related to the embedding momenta.
This prevents the direct interpretation of their quantum constraints as a
functional time Schr\"odinger equation of scalar field on a spacetime
manifold.}

In the Dirac constraint quantization, the canonical variables
$X^\pm(x)$, $\Pi_\pm(x)$, $f(x)$, $\pi_f(x)$, $m_R$, $p$
should be replaced by corresponding operators, and the constraints imposed
as restrictions on physical states $\Psi[X^\pm,f;m)$.\footnote{In 
$c=1=G$ units, the Planck constant $\hbar$ is a dimensionless number. 
One cannot make $\hbar =1$ by choosing the unit of length; 
$\hbar =1$ is a physical assumption. 
We make it from conformism, not out of inner conviction. 
Of course, it is easy to reinstate the numerical factor $\hbar$ back into 
the equations.}
The problem which needs to be overcome is that commutators of the
energy-momentum tensor operators acquire Schwinger terms \cite{BD}
which, because the constraints contain projections of the energy-momentum
tensor, enter into the commutators of the constraints as anomalies.
The imposition of constraints on the states then leads to inconsistencies.
In a previous work \cite{kuch 2} on the Dirac constraint quantization of
a massless scalar field propagating on a flat Minkowskian cylinder
${\rm I\!R}\times S^1$, we have shown how to remove the anomaly by a 
covariant (but embedding-dependent) factor ordering of the constraints.
The closing of space $S^1$ has many formal advantages (the discrete 
spectra and the removal of the infrared problem), but it is not 
appropriate in the original geometric framework of dilatonic gravity
(see the Appendix).

We now explain how the same procedure, adopted to the open space \\
\noindent 
${\rm I\!R}\times{\rm I\!R}$ boundary conditions, leads to a consistent
Dirac constraint quantization of the dilatonic model.

\subsection*{4a. Fundamental Heisenberg operators}

We want to turn the fundamental Heisenberg variables into operators acting
on a suitable function space.
Let 
\begin{equation}
\Psi=\{\Psi_n(k_1,\cdots,k_n)\}\ ,\quad n=0,1,2,\cdots
\label{eq:seq}
\end{equation}
be a sequence of complex-valued functions symmetric in their arguments
$k_1,\cdots,k_n$, with 
$\Psi_0\in
{\rm C}\llap{\vrule height 7.1pt width 1pt depth -.4pt\phantom t}$ 
being a complex number.
Define the norm of $\Psi$ by 
\begin{equation}
\label{eq:fockn}
||\Psi||^2_{{\rm Fock}}:=\Psi_0^*\Psi_0+\sum_{n=1}^\infty
\int_{-\infty}^\infty{dk_1\over |k_1|}\ \cdots
\int_{-\infty}^\infty{dk_n\over |k_n|}\ 
\Psi_n^*(k_1,\cdots,k_n)\Psi_n(k_1\cdots,k_n)\ .
\end{equation}
The finite-norm sequences $\Psi$ are elements of the familiar Fock space
${\cal F}_{\rm Fock}$.
On ${\cal F}_{\rm Fock}$, we introduce the standard annihilation
$\hat{\bf a}(k)$ and creation $\hat{\bf a}^*(k)$ operators
\begin{eqnarray}
&&\Big(\hat{\bf a}(k)\Psi\Big){}_n(k_1,\cdots,k_n)=
\sqrt{n+1}\ \Psi_{n+1}(k,k_1,\cdots,k_n)\ ,\quad\quad\quad\quad
\label{eq:acta}\\
&&\Big(\hat{\bf a}^*(k)\Psi\Big){}_n(k_1,\cdots,k_n)=\nonumber\\
&&\quad\quad\quad\quad{1\over\sqrt{n}}\sum_{i=1}^n|k|\delta(k-k_i)
\Psi_{n-1}(k_1,\cdots,k_{i-1},k_{i+1},\cdots,k_n)\ ,
\label{eq:acta*}
\end{eqnarray}
which satisfy the commutation relations
\begin{equation}
{\bf [}\ \hat{\bf a}(k),\hat{\bf a}^*(\bar k)\ {\bf ]}
=|k|\delta(k-\bar k)\ .
\label{eq:coma}
\end{equation}

We want to represent the fundamental Heisenberg operators $\hat{\bf f}(x)$
and $\hat{\bf\pi}_{{\bf f}}(x)$ on the Fock space ${\cal F}_{\rm Fock}$.
To see how this is done, it is best to introduce first the scalar field
operator
\begin{equation}
\hat f(X)={1\over \sqrt{2\pi}}{1\over\sqrt{2}}\int_{-\infty}^\infty
{dk\over|k|}\ e^{i k_\alpha X^\alpha}\ \hat{\bf a}(k)+{\rm c.c.}\ ,
\label{eq:opf}
\end{equation}
where 
\begin{equation}
\gamma_{\alpha\beta}k^\alpha k^\beta=0\ .
\end{equation}
In the double null coordinates $X^\pm$, the wave vector $k_\alpha$ has
components $k_\pm={\scriptstyle{1\over 2}}(k\mp|k|)$.
A spacelike embedding $X^\alpha(x)$ carries the canonical data
\begin{equation}
\hat f(x)=\hat f(X(x))\ ,\quad\hat\pi_f(x)=g^{{1\over 2}}(x;X]
n^\alpha(x;X]\hat f_{,\alpha}(X(x))\ ,
\label{eq:candat}
\end{equation}
where $n^\alpha(x;X]$ is the unit (future-pointing) contravariant normal to 
the embedding, and $g(x;X]$ is the determinant of the induced spatial metric.
By virtue of (\ref{eq:coma}), these data satisfy the canonical commutation
relations.
In particular, the initial embedding (\ref{eq:inem}) carries the 
fundamental Heisenberg data
\begin{eqnarray}
\hat{\bf f}(x)&=&{1\over \sqrt{2\pi}}{1\over\sqrt{2}}
\int_{-\infty}^\infty{dk\over|k|}\ 
e^{ikx}\ \hat{\bf a}(k)+{\rm c.c.}\ ,
\label{eq:heif}\\
\hat{\bf\pi}_{{\bf f}}(x)&=&{1\over \sqrt{2\pi}}{1\over\sqrt{2}}{1\over i}
\int_{-\infty}^\infty dk\ e^{ikx}\ \hat{\bf a}(k)+{\rm c.c.}\ .
\label{eq:heipi}
\end{eqnarray}
This is the desired representation of $\hat{\bf f}(x)$ and 
$\hat{\bf\pi}_{{\bf f}}(x)$ on ${\cal F}_{\rm Fock}$.

To have a space which would be able to carry a representation of the
remaining fundamental Heisenberg operators, we extend ${\cal F}_{\rm Fock}$
into a larger space ${\cal F}$ by allowing
$\Psi_n(k_1,\cdots,k_n,m;{\bf X}^\alpha]$ to be also functions of $m$
and functionals of the embeddings ${\bf X}^\alpha(x)$.
On ${\cal F}$, we represent $\hat m_R$ and $\hat{\bf X}^\alpha(x)$
by multiplication operators, and $\hat p$ and $\hat{\bf\Pi}_\alpha(x)$ by
differentiation operators:\footnote{One may want the spectrum of the mass 
operator $\hat m_R$ to be non-negative.
If so, one needs to replace the operator representation (\ref{eq:opmp})
by one appropriate for the affine group, and modify appropriately the
inner product (\ref{eq:norm}).
The details of this approach are clearly discussed by Isham \cite{Isham}.}
\begin{equation}
\hat m_R=m\times\ ,\quad\hat p=-i{\partial\over\partial m}\ ,
\label{eq:opmp}
\end{equation}
\begin{equation}
\hat{\bf X}^\alpha(x)={\bf X}^\alpha(x)\times\ ,\quad
\hat{\bf\Pi}_\alpha(x)=-i{\delta\over\delta{\bf X}^\alpha(x)}\ .
\label{eq:opem}
\end{equation}
We have thus represented all the fundamental Heisenberg operators on
${\cal F}$.

\subsection*{4b. Constraint quantization in the Heisenberg picture}

In the Dirac constraint quantization, constraints are imposed as operator
restrictions on physical states:
\begin{equation}
\hat{\bf\Pi}_\alpha(x)\ \Psi=0\ .
\label{eq:qc}
\end{equation}
In the Heisenberg picture, $\hat{\bf\Pi}_\alpha(x)$ coincide with the
embedding momenta and, as such, they are represented by variational 
derivatives (\ref{eq:opem}).
The constraint equation has a simple solution;
it implies that physical states do not depend on embeddings: 
\begin{equation}
\Psi=\{\Psi_n(k_1,\cdots,k_n,m)\}\ .
\label{eq:solpsi}
\end{equation}
This is, of course, the trademark of the Heisenberg picture:
States do not depend on time.

The space ${\cal F}_0$ of physical states can now be equipped with the
norm
\begin{equation}
||\Psi||^2=\int_{-\infty}^\infty dm\ ||\Psi(m)||_{{\rm Fock}}
\label{eq:norm}
\end{equation}
which determines the statistical predictions of the theory.${}^{11}$
Observables are to be constructed from the Schr\"odinger field variables
$\hat f(x),\hat\pi_f(x)$ on an embedding $X^\alpha(x)$.
In the Heisenberg picture, these variables are expressed as functionals
of the fundamental Heisenberg operators 
$\hat{\bf X}^\alpha(x)$, $\hat{\bf f}(x)$, $\hat{\bf\pi}_{{\bf f}}(x)$:
\begin{eqnarray}
\hat f(x)=f(x;\hat{\bf X},\hat{\bf f},\hat{\bf\pi}_{{\bf f}}]\ ,
\label{eq:fpish1}\\
\hat\pi_f(x)=\pi_f(x;\hat{\bf X},\hat{\bf f},\hat{\bf\pi}_{{\bf f}}]\ .
\label{eq:fpish2}
\end{eqnarray}
They depend explicitly on time, i.e., on the embeddings 
${\bf X}^\alpha(x)$.
The time dependence is determined by the Heisenberg equations of motion.
These are the operator versions of the classical equations
(\ref{eq:heq}), (\ref{eq:h-T}):
\begin{eqnarray}
i{\delta\hat f(x;{\bf X}]\over\delta{\bf X}^\pm(\bar x)}
&=&{\bf [}\ \hat f(x;{\bf X}],h_\pm(\bar x;{\bf X}^\pm,\hat{\bf f},
\hat{\bf\pi}_{{\bf f}}]\ {\bf ]}\\
&=&\pm{\bf [}\ \hat f(x;{\bf X}],{\bf X}^\pm{}'(\bar x)
\hat T_{\pm\pm}({\bf X}^\pm(\bar x))\ {\bf ]}\ ,
\label{eq:qheq}
\end{eqnarray}
and similarly for $\hat\pi_f(x;{\bf X}]$.
They are to be solved under the initial condition that the field operators
$\hat f(x;{\bf X}]$ and $\hat\pi_f(x;{\bf X}]$ match the fundamental
Heisenberg operators $\hat{\bf f}(x)$ and $\hat{\bf\pi}_{{\bf f}}(x)$
on the initial embedding.
The energy-momentum operator $\hat T_{\pm\pm}({\bf X}^\pm(\bar x))$ in 
(\ref{eq:qheq}) is assumed to be normal ordered in the annihilation and
creation operators (\ref{eq:acta}) and (\ref{eq:acta*}).
As one can expect of a linear field theory, the solution of the Heisenberg
equations of motion can be constructed from the scalar field operator
(\ref{eq:opf}) by differentiations, projections, and restrictions to the
embedding according to equations (\ref{eq:candat}).

We see that the Dirac constraint quantization of the dilatonic model in the
Heisenberg picture leads to the standard Fock space and scalar field 
operator of linear field theory on a flat background.
The only extra feature is the presence of a single additional degree of
freedom $m$ in the state (\ref{eq:solpsi}).
This corresponds to the primordial component of the black hole mass which
remains undetermined by the matter degrees of freedom.
Indeed, the matter degrees of freedom fix only the {\it difference} 
$(m_L-m_R)$ between the masses at the left and right infinities in accordance 
with (\ref{eq:mL}).
Because the true Hamiltonian in the canonical action (\ref{eq:Sfinal})
is equal to zero, both $m_R$ and $p$ are constants of the motion.
The same is true about the corresponding operators in quantum theory:
\begin{equation}
{\delta\hat m_R\over\delta{\bf X}^\pm(x)}=0=
{\delta\hat p\over\delta{\bf X}^\pm(x)}\ .
\label{eq:constmp}
\end{equation}
We conclude that the quantum theory in the Heisenberg picture has the
same degrees of freedom as the classical theory, and that the Heisenberg
equations of motion (which are linear) have exactly the same form as the
classical equations.

\subsection*{4c. Anomaly}

The Dirac constraint quantization of the dilatonic model is simplest when
carried out in the Heisenberg picture.
Historically, however, the Dirac procedure has always been associated with
the Schr\"odinger picture.
In this framework, its naive application to the dilatonic model leads to
inconsistencies because the constraints develop an anomaly.
To settle the question whether the Dirac procedure can be consistently
implemented in the Schr\"odinger picture, we shall trace the origins of the
anomaly and show how to remove it from the Schr\"odinger form of the
constraints.

To pass from the Heisenberg to the Schr\"odinger picture, we need to 
transform the fundamental Heisenberg operators into fundamental 
Schr\"odinger operators.
We have already seen that (\ref{eq:candat}) and (\ref{eq:opf}) solve this
problem for the field operators.
Because the embedding variables of the two pictures are the same,
(\ref{eq:h=sx}), the only remaining task is to find the Schr\"odinger
embedding momenta $\hat\Pi_\pm(x)$.
The classical equation (\ref{eq:comcon}) leads us to a natural candidate
for $\hat\Pi_\pm(x)$, namely
\begin{eqnarray}
\hat{\bar\Pi}_\pm(x)&=&\hat{\bf\Pi}_\pm(x)-
h_\pm(x;{\bf X}^\pm,\hat{\bf f},\hat{\bf\pi}_{{\bf f}}]\\
&=&\hat{\bf\Pi}_\pm(x)\mp{\bf X}^\pm{}'(x)\hat T_{\pm\pm}({\bf X}^\pm(x))\ ,
\label{eq:pibar}
\end{eqnarray}
where $\hat T_{\pm\pm}({\bf X}^\pm(x))$ is the normal ordered energy-momentum
tensor operator.
This choice does not work, however, because the operators
$\hat{\bar\Pi}_\pm(x)$ do not commute.

To see this, one evaluates first the commutators of the components
$\hat T_{\pm\pm}(X^\pm)$ at two spacetime events, $X^\pm$ and 
$\bar X{}^\pm$.
A somewhat involved but straightforward calculation based on equations
(\ref{eq:nullt}), (\ref{eq:opf}), and (\ref{eq:coma}) reveals that the
commutator differs from the classical Poisson bracket by a Schwinger
term proportional to the triply differentiated 
$\delta$-function:\footnote{The details of the calculation on the
cylindrical background ${\rm I\!R}\times S^1$ are given in
\cite{GSW} or \cite{kuch 2}.
The calculation in the open case ${\rm I\!R}\times{\rm I\!R}$ is similar
and leads to exactly the same Schwinger term.
The Casimir term present on ${\rm I\!R}\times S^1$ is absent in the open
case. The sign of the anomaly and its potential given in \cite{kuch 2}
should be corrected from $+$ to $-$. After this is done, 
to convert the signs of \cite{kuch 2}
into those in the present paper, one needs to keep track of the switch
from the $T^{\pm} := T\pm Z$ variables used in \cite{kuch 2} to the 
$X^{\pm}$ variables of equation (\ref{eq:9}).}
\begin{eqnarray}
{1\over i}{\bf [}\ \hat T_{\pm\pm}(X^\pm),
\hat T_{\pm\pm}(\bar X{}^\pm)\ {\bf ]}&=&
\pm\hat T_{\pm\pm}(X^\pm)\delta_{,\pm}(X^\pm-\bar X{}^\pm)
-(X^\pm\leftrightarrow \bar X{}^\pm)\nonumber\\
&&\mp{1\over 12}{1\over 2\pi}\delta_{,\pm\pm\pm}(X^\pm-\bar X{}^\pm)\ ,
\label{eq:schw1}
\end{eqnarray}
while
\begin{equation}
{\bf [}\ \hat T_{++}(X^+),\hat T_{--}(\bar X{}^-)\ {\bf ]}=0\ .
\label{eq:schw2}
\end{equation}
The Schwinger term leads then to the anomaly
\begin{equation}
F_{\pm\pm}(x,\bar x;{\bf X}^\pm]=\pm{1\over 12}{1\over 2\pi}
\partial_x\Bigg(\left({\bf X}^\pm{}'(x)\right)^{-1}
\partial_x\bigg(\left({\bf X}^\pm{}'(x)\right)^{-1}
\partial_x\delta(x,\bar x)\bigg)\Bigg)
\label{eq:an}
\end{equation}
in the commutator of (\ref{eq:pibar}):
\begin{equation}
{1\over i}{\bf [}\ \hat{\bar\Pi}_\pm(x),
\hat{\bar\Pi}_\pm(\bar x)\ {\bf ]}=
-F_{\pm\pm}(x,\bar x;{\bf X}^\pm]\ .
\label{eq:compibar}
\end{equation}
The details of these as well as of the following calculations can be
found in \cite{kuch 2}.${}^{12}$

To summarize, the operators $\hat{\bar\Pi}_\pm(x)$ commute with the 
Schr\"odinger fields $\hat f(x)$ and $\hat\pi_f(x)$---this is the content
of the Heisenberg equations of motion (\ref{eq:qheq})---and they also have 
the correct commutators with the embeddings:
\begin{equation}
{1\over i}{\bf [}\ \hat X^\pm(x),\hat{\bar\Pi}_\pm(\bar x)\ {\bf ]}
=\delta(x,\bar x)\ .
\label{eq:comxpibar}
\end{equation}
However, they do not commute among themselves.
This prevents us from identifying them with the Schr\"odinger embedding
momenta.

\subsection*{4d. Removing the anomaly}

Let us show how to amend $\hat{\bar\Pi}_\pm(x)$ into commuting operators
which retain the correct commutators with the rest of the fundamental
Schr\"odinger variables.
The clue is provided by the Jacobi identity of the commutator
(\ref{eq:compibar}).
Because the anomaly depends only on the embedding variables, we 
conclude that
\begin{equation}
{\delta F_{\alpha\beta}(x,\bar x;{\bf X}]\over\delta{\bf X}^\gamma
(\bar{\bar x})}
+{\rm cyclic\ permutations}\ (\alpha x,\beta\bar x,\gamma\bar{\bar x})=0\ .
\label{eq:anclose}
\end{equation}
This means that the anomaly is a closed 2-form on the space of 
embeddings.
More than that, the anomaly is exact:
There exists a potential $A_\alpha(x;{\bf X}]$
(we shall call it the anomaly potential) whose exterior derivative
generates $F_{\alpha\beta}(x,\bar x;{\bf X}]$:
\begin{equation}
F_{\alpha\beta}(x,\bar x;{\bf X}]=
{\delta A_\beta (\bar x,{\bf X}]\over\delta{\bf X}^\alpha(x)}-
{\delta A_\alpha(x,{\bf X}]\over\delta{\bf X}^\beta(\bar x)}\ .
\label{eq:anex}
\end{equation}
Of course, $A_\alpha(x;{\bf X}]$ is determined by (\ref{eq:anex}) only
up to a functional gradient.

It is easy to check that
\begin{equation}
A_\pm(x;{\bf X}^\pm]=\pm{1\over 24}{1\over 2\pi}
\bigg(\left({\bf X}^\pm{}'(x)\right)^{-1}\bigg)^{\prime\prime}
\label{eq:ana}
\end{equation}
is one anomaly potential;
indeed, the exterior derivative (\ref{eq:anex}) of (\ref{eq:ana})
gives the anomaly (\ref{eq:an}).
Unfortunately, (\ref{eq:ana}) does not transform as a scalar density
under spatial diffeomorphisms.
To improve this, we can gauge (\ref{eq:ana}) by a functional gradient
into a new potential
\begin{eqnarray}
A_\pm(x;{\bf X}]&=&{1\over 24}{1\over 2\pi}
\Bigg(\left({\bf X}^\pm{}'(x)\right)^{-1}\Bigg({{\bf X}^-{}''(x)
\over{\bf X}^-{}'(x)}-{{\bf X}^+{}''(x)\over{\bf X}^+{}'(x)}\Bigg)
\Bigg)^\prime
\nonumber\\
&=&{1\over 12}{1\over 2\pi}\bigg(\left({\bf X}^\pm{}'(x)\right)^{-1}
g^{{1\over 2}}(x;{\bf X}] K(x;{\bf X}]\bigg)^\prime\ ,
\label{eq:a}
\end{eqnarray}
which is a scalar density and generates the same anomaly.
As indicated, the potential (\ref{eq:a}) is simply related to the mean
extrinsic curvature $K$ of the embedding.

Our old operators $\hat{\bar\Pi}_\pm(x)$ have correct commutation
relations with the Schr\"odinger fields $\hat f(x)$, $\hat\pi_f(x)$ and
with $\hat X^\pm(x)$, but they do not commute among themselves.
By subtracting from them the anomaly potential, we change them into
commuting operators
\begin{equation}
\hat\Pi_\pm(x):=\hat{\bar\Pi}_\pm(x)-A_\pm(x;{\bf X}]\ ,
\label{eq:pi}
\end{equation}
which retain the correct commutation relations with $\hat f(x)$,
$\hat\pi_f(x)$, and $\hat X^\pm(x)$.
These we can identify with the Schr\"odinger embedding momenta.

\subsection*{4e. Imposing constraints in the Schr\"odinger picture}

At this point, we can express the constraints in terms of the
fundamental Schr\"odinger variables:
\begin{equation}
\hat{\bf\Pi}_\pm(x)=\hat\Pi_\pm(x)+h_\pm(x;X^\pm,\hat f,\hat\pi_f]
+A_\pm(x;X]\approx 0\ .
\label{eq:schrcon}
\end{equation}
To find an explicit form of  $h_\pm(x;X^\pm,\hat f,\hat\pi_f]$, we split
the spacetime field operators $\hat f_{,\pm}(X^\pm)$ into their
positive-frequency (the Heisenberg annihilator $\hat{\bf a}(k)$)
and negative-frequency (the Heisenberg creator $\hat{\bf a}^*(k)$)
parts:
\begin{equation}
\hat f_{,\pm}(X^\pm)={}_{(+)}\hat f_{,\pm}(X^\pm)+
{}_{(-)}\hat f_{,\pm}(X^\pm)\ .
\label{fsplit}
\end{equation}
This is achieved by the positive- and negative-frequency parts
${}_{(\pm)}\delta$ of the $\delta$-function
\begin{equation}
{}_{(\pm)}\delta(X):={1\over 2\pi}\int_0^\infty dk\ e^{\mp ikX}\ ,
\label{eq:pndelta}
\end{equation}
which act as kernels of integral operators
\begin{equation}
{}_{(\pm)}\hat f_{,\pm}(X^\pm)=\int_{-\infty}^\infty d\bar X{}^\pm\ 
{}_{(\pm)}\delta(X^\pm-\bar X{}^\pm)\hat f_{,\pm}(\bar X{}^\pm)\ .
\label{eq:fsplex}
\end{equation}
This decomposition allows us to perform the Heisenberg normal ordering 
of the energy-momentum tensor by the ordering kernel
\begin{eqnarray}
\lefteqn{{\cal N}(X^\pm;\bar X{}^\pm,\bar{\bar X}{}^\pm)=}
\nonumber\\
&&{}_{(+)}\delta(X^\pm-\bar X{}^\pm)\ {}_{(+)}\delta(X^\pm-\bar{\bar X}{}^\pm)
+{}_{(-)}\delta(X^\pm-\bar X{}^\pm)\ {}_{(-)}\delta(X^\pm-\bar{\bar X}{}^\pm)
\nonumber\\
&&+{}_{(-)}\delta(X^\pm-\bar X{}^\pm)\ {}_{(+)}\delta(X^\pm-\bar{\bar X}{}^\pm)
+{}_{(-)}\delta(X^\pm-\bar{\bar X}{}^\pm){}_{(+)}\delta(X^\pm-\bar X{}^\pm)
\nonumber\\ 
\label{eq:orn}
\end{eqnarray}
in an integral operator
\begin{equation}
\hat T_{\pm\pm}(X^\pm)=
\int_{-\infty}^\infty d\bar X{}^\pm\int_{-\infty}^\infty d\bar{\bar X}{}^\pm\ 
{\cal N}(X^\pm;\bar X{}^\pm,\bar{\bar X}{}^\pm)
\hat f_{,\pm}(\bar X{}^\pm)\hat f_{,\pm}(\bar{\bar X}{}^\pm)\ .
\label{eq:tor}
\end{equation}
The connection (\ref{eq:fcan}) between the field operators 
$\hat f_{,\pm}(x)$ and the fundamental Schr\"odinger operators then finishes 
our task:
\begin{eqnarray}
\lefteqn{h_\pm(x;X^\pm,\hat f,\hat\pi_f]=\pm{1\over 4}X^\pm{}'(x)
\int_{-\infty}^\infty d\bar x\int_{-\infty}^\infty d\bar{\bar x}\ \times}
\nonumber\\
&&\times{\cal N}(X^\pm(x);X^\pm({\bar x}),X^\pm(\bar{\bar x}))
(\hat f'(\bar x)\pm\hat\pi_f(\bar x))
(\hat f'(\bar{\bar x})\pm\hat\pi_f(\bar{\bar x}))\ .
\label{schrh}
\end{eqnarray}

The fundamental Schr\"odinger operators satisfy the appropriate
commutation relations:
\begin{eqnarray}
{1\over i}{\bf [}\ \hat X^\pm(x),\hat\Pi_\pm(\bar x)\ {\bf ]}
&=&\delta(x,\bar x)\ ,
\label{eq:schrcom1}\\
{1\over i}{\bf [}\ \hat f(x),\hat\pi_f(\bar x)\ {\bf ]}
&=&\delta(x,\bar x)\ ,
\label{eq:schrcom2}
\end{eqnarray}
with all other commutators vanishing.
In the Schr\"odinger picture, $\hat X^\pm(x)$ and $\hat\Pi_\pm(x)$
are represented as multiplication and differentiation operators
\begin{equation}
\hat X^\pm(x)=X^\pm(x)\times\ ,\quad
\hat\Pi_\pm(x)=-i{\delta\over\delta X^\pm(x)}
\label{eq:semop}
\end{equation}
and the quantum constraint (\ref{eq:qc}) yields the functional
Schr\"odinger equation
\begin{equation}
i{\delta\Psi[X]\over\delta X^\pm(x)}=\left(h_\pm(x;X^\pm,\hat f,\hat\pi_f]
+A_\pm(x;X]\right)\ \Psi[X]=0\ .
\label{eq:schreq}
\end{equation}
This determines the dependence of the Schr\"odinger state $\Psi[X]$
on the embeddings.
If one also decides to work in the Schr\"odinger {\it representation},
the field operators get represented by multiplication and differentiation
operators, and (\ref{eq:schreq}) becomes an equation for the state 
functional $\Psi[X,f]$.
The anomaly potential ensures the functional integrability of 
(\ref{eq:schreq}).

We thus see that the Dirac constraint quantization in the Schr\"odinger 
picture, though somewhat subtle to formulate, is consistent and equivalent 
to the more straightforward Heisenberg picture quantization.

\section*{5. How to quantize physical geometry?}

We found a canonical transformation which reduced the constraints of the
dilatonic model to those of a parametrized field theory on a flat background.
Any reference to the original physical geometry disappeared from the
description of the system.  
However, the interesting questions in dilatonic gravity are precisely those
which are concerned with the physical spacetime.
We must show how to pose such questions in the framework based on the flat
background canonical variables.

We know that the physical metric $\bar\gamma_{\alpha\beta}$ is related to
the auxiliary flat metric $\gamma_{\alpha\beta}$ by the dilaton factor
$y^{-1}$, (\ref{eq:physical_metric}).
In the null coordinates $X^\pm$, the physical interval $d\bar s$ is 
given by the formula
\begin{equation}
d\bar s^2={dX^+ dX^-\over y(X)}\ .
\label{eq:sbar}
\end{equation}
In the canonical theory, the dilaton is connected to the new canonical
variables $(X^\pm(x),\Pi_\pm(x);m_R,p)$ by the canonical transformation
(\ref{eq:y(X)}).
On the constraint surface (\ref{eq:comcon}), (\ref{eq:h-T}), the dilaton
can be expressed as a functional of the energy-momentum tensor:
\begin{eqnarray}
y(X)&=&\kappa^2X^+ X^- 
\nonumber\\
&&-\int_\infty^{X^-}d\bar X{}^-\int_\infty^{{\bar X}{}^-}d\bar{\bar X}{}^-\  
T_{--}(\bar{\bar X}{}^-) 
-\int_{-\infty}^{X^+}d\bar X{}^+\int_{-\infty}^{{\bar X}{}^+}
d\bar{\bar X}{}^+\ 
T_{++}(\bar{\bar X}{}^+)
\nonumber\\
&&-\int_{-\infty}^\infty dX^+\ X^+T_{++}(X^+) + {m_R\over \kappa}\ .
\label{eq:dil}
\end{eqnarray}
This brings us back to the spacetime solution (\ref{eq:y1})-(\ref{eq:y2}).

By quantizing the dilatonic model in the Heisenberg picture, we turn
$\hat y(X)$ into an operator on the Fock space ${\cal F}_0$ of 
physical states (\ref{eq:solpsi}) with the norm (\ref{eq:norm}).
This turns the network of physical intervals (\ref{eq:sbar}) into 
operators.
The operator version of equation (\ref{eq:sbar}) is thus the starting 
point of discussions about quantum geometry.

To make sense of the operator version of equation (\ref{eq:sbar}) is not 
entirely straightforward.
First, in the old action, the dilaton field $y(x)$ is one of the canonical
coordinates, and hence it commutes at any two points on a spacelike surface.
When expressed as a functional (\ref{eq:dil}) of the scalar field data, this 
is no longer necessarily true.
This poses questions about simultaneous measurability of different pieces 
of the quantum geometry.
Second, one must decide what is the correct factor ordering of the field 
operators in the energy-momentum tensor in (\ref{eq:dil}).
Normal ordering is a natural candidate, but our previous discussion 
(section 4c) raises the possibility that other options may be more 
appropriate.
Third, after the factor ordering decision is made, the quantum geometry
should be defined by spectral analysis. 
Finally, the classical dilaton field is required to be positive, to ensure 
the right signature $(-,+)$ of the physical metric. 
In quantum theory it is quite difficult to maintain the positivity condition. 
Indeed, refraining from doing so opens the possibility of evolving the 
physical geometry through what classically would appear to be a singularity. 
These and other problems must be settled before one can meaningfully speak 
about quantum physical geometry. 
We intend to address them in a future paper.

Notwithstanding that the detailed resolution of issues connected with 
the Hawking effect will depend on how we settle the problems mentioned 
above, we stress that the quantum theory in section 4 is a standard unitary 
quantum field theory on a Fock space; i.e., that we do not encounter any 
loss of unitarity under evolution.

\section*{Acknowledgments}
We want to thank Ji\v{r}\'{\i} Bi\v{c}\'{a}k and Laszlo Gergely for 
interactions in early stages of our research, and Roman Jackiw and 
Aleksandar Mikovi\'{c} for discussing their work with us.
We would also like to thank Sukanta Bose and Yoav Peleg  for helpful
discussions.

This work was supported in part by NSF grants PHY-9507719 to the University 
of Utah, PHY-9507740 to the University of Wisconsin-Milwaukee (JDR),
and U.S.-Czech Republic Science and Technology Program grant 92067 (KVK).

\section*{Appendix}
To see the problems of putting the CGHS model on  $ {\rm I\!R}\times S^1$ 
in its simplest setting, inspect the vacuum solution
\begin{equation}
y(X)= \kappa^2 X^+ X^- + {m\over\kappa}
\label{eq:A1}
\end{equation}
for the dilaton field. 
This describes a primordial black hole of mass $m$ with the physical metric 
$\bar {\gamma}_{\alpha \beta} = y^{-1}(X)\gamma_{\alpha \beta}$ 
(see \cite{str}). 

Try to put this solution on a Minkowskian cylinder ${\rm I\!R}\times S^1$ 
by identifying the points with Cartesian coordinates
$ X^{\alpha}_{(1)}=(T, Z- \xi )$ and $X^{\alpha}_{(2)}=(T, Z + \xi )$;
i.e., with the double null coordinates differing by a translation:
\begin{equation}
X^{\pm}_{(2)}=X^{\pm}_{(1)} + 2\xi\ ,\quad\quad \xi\in (0,\infty)\ .
\label{eq:A2}
\end{equation}

The dilaton $y(X)$, the induced metric $g_{TT}(T)= -y^{-1}$,  and the 
scalar curvature $R[\bar{\gamma}]= 4m \kappa y^{-1}$ are continuous 
across the seam. 
However, the (mean) extrinsic curvature of the seam, as embedded in the 
physical geometry on its two sides, suffers a jump: 
\begin{equation}
K(\xi) - K(-\xi) = 2\kappa\cdot\kappa\xi\cdot y^{-{1\over 2}}\ .
\end{equation}
This indicates the presence of an unphysical sheet of matter. 
We would encounter the same situation if we tried to identify the left 
and the right openings of the Einstein-Rosen bridge at the same finite 
value of the spatial Kruskal coordinate (and hence the same value of the 
area coordinate $r$, which is analogous to the dilaton) in the 
Schwarzschild black hole. 
Our argument  can easily be generalized to the dilatonic black hole 
(\ref{eq:y1})-(\ref{eq:y2}) in the presence of a scalar field. 
This straightforward attempt of putting the dilatonic black hole on a 
Minkowskian cylinder thus fails.

The correct way of putting the vacuum black hole solution 
(\ref{eq:A1}) on ${\rm I\!R}\times S^1$ is to cut a wedge from the 
dynamical region $X^+>0$, $X^-<0 $ and wrap it into a cone.\footnote{The 
same construction can also be done in the past dynamical region
$X^+<0$, $X^->0$.}  
Take a straight timelike line 
\begin{equation}
X^{\alpha}(t)= t^\alpha t 
\label{eq:A5}
\end{equation}
in the Minkowskian plane parametrized by the Minkowskian proper time $t>0$,
with $t^\alpha$ being a constant, future-pointing, unit tangent vector with 
respect to the flat Minkowski metric.
The vector field $k^\alpha$, having components
\begin{equation}
k^{\pm}= (\kappa X^+,-\kappa X^-)\ ,
\end{equation}
is perpendicular to the radius vector $X^{\pm}= (X^+, X^- )$, and it is a 
Killing vector field of the physical metric. 
Because $k^\alpha$ is perpendicular to $t^\alpha$, the extrinsic curvature 
of the line (\ref{eq:A5}), as embedded in the physical spacetime, vanishes. 
Its induced metric is $g_{tt}= -(-\kappa^2 t^2+\kappa^{-1}m)^{-1}$.

Now take any two lines, with tangent vectors $t^{\alpha}_{(1)}$
and $t^{\alpha}_{(2)}$ oriented in the clockwise direction with respect 
to one another, and identify their points labeled by the same $t$.  
Instead of the Minkowskian cylinder (\ref{eq:A2}), we get a Minkowskian 
cone 
\begin{equation}
X^{+}_{(2)}= X^{+}_{(1)} e^{\eta}\ , \quad
X^{-}_{(2)}= X^{-}_{(1)} e^{-\eta}\ ,
\end{equation}
where the monodromy parameter $\eta$ is the hyperbolic angle between the 
two lines in the Minkowskian plane. 
Because the physical metric is conformally related to the flat Minkowski 
metric, it is also the hyperbolic angle between the lines in the physical 
space. 
Both the induced metric and the extrinsic curvature match at the seam. 
The transformation equations
\begin{equation}
X^{\pm} = \pm\kappa^{-1}\sqrt{m\over \kappa}
\exp\left({\eta \over 2\pi}(\chi\pm\phi) \pm \eta_0\right)\ ,
\label{eq:A8}
\end{equation}
where $\eta_0 $ is the hyperbolic angle of the midline between  
$t^{\alpha}_{(1)}$ and $t^{\alpha}_{(2)}$, brings us to dimensionless
coordinates $\chi\in(-\infty ,0)$, $\phi\in(-\pi,\pi)$ and the physical 
line element
\begin{equation}
d{\bar{s}}^2 = \bigg({\eta \over 2\pi \kappa}\bigg)^2
\left(e^{-{\eta\chi\over\pi}} - 1\right)^{-1}(-d\chi^2 + d\phi^2)\ .
\label{eq:A9}
\end{equation}
The vector field $\partial\over{\partial\phi}$ is a Killing field of the
physical metric.

The resulting ${\rm I\!R}\times S^1$ spacetime covers only a segment of the 
interior of the black hole, and it is geodesically incomplete. 
The counterpart of this construction for the Schwarzschild black hole 
is the Kantowski-Sachs universe. 
Again, one can generalize this procedure to the dilatonic black hole 
(\ref{eq:y1})-(\ref{eq:y2}) produced by a scalar source \cite{mik 4}.
For possible ${\rm I\!R}\times S^1$ identifications in different dilatonic 
theories, see \cite{strob}. 

Thus, we see that by putting the CGHS model on ${\rm I\!R}\times S^1$,  
we lose the physical picture of the black hole formation by the gravitational 
collapse of the matter field.

\end{document}